# Manipulating infrared photons using plasmons in transparent graphene superlattices

Hugen Yan, Xuesong Li, Bhupesh Chandra, George Tulevski, Yanqing Wu,
Marcus Freitag, Wenjuan Zhu, Phaedon Avouris[*], and Fengnian Xia[*]

**Abstract.** Superlattices are artificial periodic nanostructures which can control the flow of electrons[1, 2]. Their operation typically relies on the periodic modulation of the electric potential in the direction of electron wave propagation. Here we demonstrate transparent graphene superlattices which can manipulate infrared photons utilizing the collective oscillations of carriers, i.e., plasmons[3-10] of the ensemble of multiple graphene layers. The superlattice is formed by depositing alternating wafer-scale graphene sheets and thin insulating layers, followed by patterning them all together into 3-dimensional photonic-crystal-like structures. We demonstrate experimentally that the collective oscillation of Dirac fermions in such graphene superlattices is unambiguously nonclassical: compared to doping single layer graphene, distributing carriers into multiple graphene layers strongly enhances the plasmonic resonance frequency and magnitude, which is fundamentally different from that in a conventional semiconductor superlattice[5, 6, 10]. This property allows us to construct widely tunable far-infrared notch filters with 8.2 dB rejection ratio and terahertz linear polarizers with 9.5 dB extinction ratio, using a superlattice with merely five graphene atomic layers. Moreover, an unpatterned superlattice shields up to 97.5% of the electromagnetic radiations below 1.2 terahertz. This demonstration also opens an avenue for the realization of other transparent mid- and far-infrared photonic devices such as detectors, modulators, and 3-dimensional meta-material systems[11-13].

*Email: fxia@us.ibm.com (F.X.) and avouris@us.ibm.com (P.A.)

Despite of being atomically thin, graphene interacts with light strongly within a wide wavelength range and consequently attracts significant attention[14-17]. It exhibits particularly strong potential in far-infrared and terahertz optoelectronics due to its high carrier mobility and conductivity[7, 18-21]. The light-graphene interaction in this frequency range can be described by the Drude model using the dynamical conductivity $\sigma(\omega) = \frac{iD}{\pi(\omega + i\Gamma)}$, where $D$ is the Drude weight and $\Gamma$ is the carrier scattering width. The Drude weight $D$, for graphene with linear energy bands, is $\frac{V_F e^2}{\hbar}\sqrt{\pi|n|}$, where $V_F$ is the Fermi velocity and $n$ is the carrier density[22, 23]. In this letter, we report our study of the interaction of the far-infrared light and wafer-scale graphene superlattices. We show experimentally that the collective oscillation of carriers in graphene superlattices is distinctly different from that in regular semiconductor superlattices, although such plasmonic excitations arise from the same long-ranged Coulomb interaction in both cases. Furthermore, several far-infrared photonic devices suitable for realistic applications are demonstrated based on graphene superlattices.

The fabrication processes of the graphene superlattice are shown in Fig. 1a. It consists of multiple layer deposition steps and a single lithography step, as discussed in the Methods Summary. Figure 1b shows a scanning electron micrograph of a patterned graphene superlattice, consisting of stacks of circular graphene disks arranged in a triangular lattice. Here, $d$ is the diameter of the graphene disks and $a$ is the lattice constant[24]. The properties of the as-prepared graphene superlattices before lithographical patterning are discussed first, followed by a detailed discussion of the plasmons in patterned graphene superlattices. The inset of Fig. 1c is a photograph of an as-prepared transparent superlattice consisting of 5 graphene layers on a quartz substrate. The measured relative light transmission from ultraviolet to mid-infrared, $T/T_0$, where $T$ and $T_0$ are the transmission through the quartz sample with and without the superlattice, respectively, is shown in Fig. 1c. From 3 μm to 1.25 μm, the relative transmission is close to unity, due to strong Pauli-blocking effect[15]. The transmission decreases to around 90% at 870 nm and remains approximately constant in the entire visible wavelength range, as in the case of previous observations on exfoliated multilayer-graphene samples[14].

We first characterized the response in the far-infrared of wafer scale single layer graphene. The characterization processes are discussed in the Methods Summary. The extinction in the transmission, $1-T/T_0$, is related to the dynamical conductivity, $\sigma(\omega)$[25]:

$$1 - T/T_0 = 1 - \frac{1}{\left|1 + Z_0 \sigma(\omega)/(1+n_s)\right|^2} \qquad (1)$$

where $Z_0$ is the vacuum impedance, $\omega$ is the frequency of the light, and $n_s$ is the refractive index of the quartz substrate. The red and grey squares represent $1-T/T_0$ measured using single layer graphene on the quartz substrate with and without the polymer buffer layer underneath it, respectively. The corresponding solid curves are fitted results using Eq. (1) and the Drude dynamical conductivity, using $D$ and $\Gamma$ as two fitting parameters. The doping of graphene in both cases is between 8 to $9 \times 10^{12}$ cm$^{-2}$. The scattering width, $\Gamma$, increases from 72 cm$^{-1}$ for graphene on polymer buffer layer to 92 cm$^{-1}$ on quartz, due to the reduction of mobility induced by the quartz substrate[26]. The doping of graphene increases to around $2.5 \times 10^{13}$ cm$^{-2}$ after exposure to nitric acid vapor, as can be estimated by the position of Pauli-blocking shown in Fig. 1c. Detailed carrier density estimation procedures are presented in the Supplementary Information. The scattering width is reduced to 52 cm$^{-1}$ after doping, corresponding to an effective scattering time of 0.1 picosecond.

The extinction spectra, $1-T/T_0$, measured on as-prepared, transparent superlattices with 1, 2, 3, 4, and 5 graphene layers are shown in Fig. 2b. A schematic of such layered graphene superlattice is shown in the inset of Fig. 2c. Since the buffer layers are thin (~20 nm) and non-conducting, their contribution to the total conductivity can be ignored. Consequently, the total dynamical conductivity is the sum of the conductivities of the individual graphene layers. If the scattering widths in all graphene layers are similar, the total conductivity, $\sigma_{Total}$, is approximately $\sigma_{Total} = \frac{iD_{Total}}{\pi(\omega+i\Gamma)}$, where $D_{Totlal}$ is the sum of the Drude weights of the individual graphene layers. The solid lines in Fig. 2b are fitted curves using $D_{Totlal}$ and $\Gamma$ as the only fitting parameters based on equation (1). The fitting results are shown in Fig. 2c. The Drude weight does increase proportionally to the layer number, indicating the uniform doping in the superlattice. Moreover, the fitted scattering width remains within 50 to 60 cm$^{-1}$, confirming the uniform graphene quality. An unpatterned superlattice with 5 graphene layers results in an extinction in transmission of up to 97.5% at frequencies below 1.2 THz as shown in Fig. 2b, making it an effective optically transparent and flexible microwave and terahertz shielding material. Moreover, the overall surface electrical resistance of this graphene superlattice inferred from the total dynamical conductivity is only 25 Ω/□, which compares favorably with that of commercial indium tin oxide with a thickness of around 100 nm.

In order to realize frequency selectivity and other photonic functions beyond broadband electromagnetic wave shielding, we introduce plasmonic resonances in a graphene superlattice by lithographically patterning it into microdisks arranged in a triangular lattice as shown in Fig. 1b. Similar plasmonic structures in conventional two dimensional electron gases (2-DEG) have previously been studied[4, 5, 10, 27]. Plasmons in graphene have also been investigated[6-9, 18-20, 28]. Very recently, Ju et al. studied the localized plasmons formed in single layer graphene micro-ribbons[7]. In contrast to graphene micro-ribbons, light-plasmon coupling in graphene disks has no polarization dependence, which is desirable in applications such as filtering and detection. For the graphene disk array, neglecting the lateral disk-disk interaction and in a quasi-electrostatic approximation, the average sheet optical conductivity in the far infrared regime is[27]

$$\sigma(\omega) = i\frac{fD}{\pi} \frac{\omega}{(\omega^2 - \omega_p^2) + i\Gamma_p \omega} \quad (2)$$

where $\omega$ is the frequency, $f$ is the filling factor (graphene area over total area), $D$ is the Drude weight, and $\Gamma_P$ is the plasmon resonance width. The resonance frequency is[27, 29]

$$\omega_p = \sqrt{\frac{3D}{8\varepsilon_m \varepsilon_0 d}} \quad (3)$$

where $\varepsilon_m$ is the media dielectric constant, $\varepsilon_0$ is the vacuum permittivity and $d$ is the diameter of the graphene disks. For single layer graphene, Eq. (3) indicates that $\omega_p \propto n^{1/4}$, which is in contrast to that obtained for a conventional 2-DEG ($\omega_p \propto n^{1/2}$)[4, 27]. The optical conductivity of such plasmonic graphene superlattices can still be inferred using Eq. (1) based on the measured extinction spectra.

Figure 3a shows the extinction spectra, $1-T/T_0$, for superlattice micro-disk arrays with 1, 2, and 5 graphene layers. In all three superlattices, the disk diameter $d$ and the triangular lattice constant $a$ are 4.4 and 9 microns, respectively. From the Drude conductivity response of the unpatterned area of the superlattice, we conclude that in these 3 samples, each graphene layer has almost identical carrier density. Two prominent features are observed in Fig. 3a. First, the resonance frequency up-shifts with increasing number of layers; second, the peak intensity increases significantly with the graphene layer number. The latter can be easily understood as a consequence of the overall conductivity enhancement. We fit the spectra according to Eqs. (1) and (2) and solid lines are fitting curves. For the single layer graphene disks, the resonance frequency ($\omega_p$) and resonance width ($\Gamma_P$) from the fitting are 114 cm$^{-1}$ and 65 cm$^{-1}$, respectively (Fitting procedures are presented in Methods Summary). The significant up-shift of the resonance frequency in the graphene superlattice is due to the strong Coulomb interaction of the adjacent layers. The in-phase collective motion of the carriers among the layers results in stronger restoring force through dipole-dipole coupling[30]. In our case, the graphene layer spacing $l$ (~20 nm) is much smaller than the disk diameter $d$ (4.4 μm), and the strong coupling condition is well-satisfied[5, 31]. In conventional semiconductor superlattices with parabolic electronic energy dispersion, when two 2-DEG layers with the same carrier density $n$ are close enough, the collective oscillation is expected to be equivalent to that in a single 2-DEG layer with carrier density of $2n$, which was confirmed experimentally[10]. However, our results here reveal a striking difference between graphene and conventional 2-DEG superlattices.

The grey squares in Fig. 3b represent the normalized plasmonic resonance frequencies in superlattice versus the total normalized carrier density. In the superlattice, the enhancement of total carrier density, $n_{Total}$, is a result of the increase in layer number ($n_0$ is the carrier concentration in an individual graphene layer); clearly it exhibits a $n^{1/2}$ dependence. For comparison, in the same figure, we also show results obtained using a micro-disk array with the same $d$ and $a$ made from a single layer graphene at different chemical doping levels. The details on doping level determination and its control are discussed in Supplementary Information. The resonance follows the $n^{1/4}$ rule well (red squares in Fig. 3b), as also found by Ju et al[7]. Our study indicates that distributing Dirac Fermions in a single layer of graphene disks into multiple layers of closely stacked graphene disks can drastically increase the plasmonic resonance frequency. This can not be understood from classical plasmon theory since the Coulomb interaction depends solely on the total carrier density in the strong coupling regime. The origin of this peculiar behavior is the massless nature of the Dirac Fermions in graphene, in which the classical definition of effective mass does not apply. This can only be understood quantum mechanically, since the plasmon resonance of massless Dirac Fermions is proportional to $\hbar^{-1/2}$, as pointed out by Das Sarma et al[6, 31]. Alternatively, a carrier density dependent effective "plasmon mass"[8], $m_p = \hbar\sqrt{\pi|n|}/V_f$, can also be introduced to explain this unusual behavior, where $n$ is the carrier density in an individual graphene layer. The frequency change due to the carrier redistribution can then be regarded as a result of the discontinuous change of the plasmon mass, even though the Coulomb restoring force is still the same. Compared to conventional 2-DEG, the single layer graphene plasmonic resonance has weaker ($n^{1/4}$ versus $n^{1/2}$) carrier density dependence. This significantly limits the frequency tunability of plasmonic resonance in a single layer graphene. Moreover, the Drude weight in graphene also has weak ($n^{1/2}$) carrier density dependence. This implies that increasing carrier density does not efficiently enhance the plasmonic resonance magnitude in single layer graphene. The red squares in Fig. 3c represent the normalized dynamical conductivity measured at plasmonic resonance versus the normalized total carrier density in single layer graphene. Indeed, the carrier density dependence is much weaker than that in the graphene superlattice as shown in the grey squares in Fig. 3c, in which conductivity addition is observed and the peak conductivity at plasmonic resonance increases linearly with the layer number. Hence, using a graphene superlattice structure, one can tune both the plasmonic resonance frequency and magnitude effectively, which can potentially lead to practical photonic devices covering both mid- and far-infrared. However, conductivity addition only applies to the localized plasmons at strong coupling limit as shown here, and general plasmons in graphene superlattice need to be treated more rigorously using quantum mechanics.

The peak transmission extinction is about 50% in a superlattice with 5 graphene layers, as shown in Fig. 3a, and higher extinction can be achieved readily by enhancing the filling factor, *f*, as indicated by Eq. (2). Figure 4a shows the extinction spectra for two different disk arrays with identical *d* (4.4 μm) but different lattice constant *a* (4.8 and 9 μm) fabricated using single layer graphene. As expected, the densely packed single layer graphene disk-array exhibits much higher peak extinction. However, it also shows lower plasmonic resonance frequency. This is due to Coulomb interactions among disks within the same graphene layer, which screen the electric fields and weaken the restoring force as illustrated in the inset of Fig. 4a (see Supplementary Information for detailed discussions on disk-disk interaction within the same graphene layer)[30]. Plasmonic notch filters are constructed using superlattices with 5 graphene layers patterned into microdisks with high filling factor. Figure 4b shows the extinction spectra of such filters with different $d$. The lattice constant $a$, in all filters, is designed to be only 400 nm larger than $d$. Using such densely packed micro-disk arrays, we achieve peak extinction as high as 85%, corresponding to a notch filter with an extinction ratio of around 8.2 dB. The peak extinction frequency can be engineered by adjusting the $d$, as shown in Fig. 4b. We did not observe any appreciable polarization dependence in these notch filters.

With a 5-layer graphene superlattice, we also made polarizers by etching it into micro-ribbon arrays. Fig. 4c shows the extinction spectra for the light parallel and perpendicular to the ribbon axis. The ribbon width ($W$) is 2 μm and the spacing between the ribbons ($S$) is 500 nm. The plasmonic resonance peak, which only appears when the electric field is perpendicular to the ribbon axis[7], is beyond the measurement range set by the commercial linear polarizer used in the experiment. We plot the extinction ($1-T/T_0$) measured at 40cm$^{-1}$ as a function of light polarization in the upper left inset of Fig.4c. The extinction of undesirable polarization, defined as $10\left|\lg(T/T_0)_{\theta=0^0}\right|$ at 40cm$^{-1}$ reaches 9.5 dB.

Compared with conventional semiconductor superlattices, simultaneously high carrier density and high carrier mobility in graphene superlattices leads to demonstration of practical photonic devices using merely five atomically graphene layers. Moreover, room temperature operation, tunability, and scalability to larger size also make them technologically relevant.

The introduction of the superlattice structure may have important implications for graphene research. It first provides an ideal platform for the scientific investigation of interactions of massless Dirac Fermions. Using a simple 3-dimensional graphene superlattice, we show experimentally for the first time that the collective oscillation of Dirac fermions is unambiguously quantum mechanical. Furthermore, three far-infrared photonic devices transparent in the visible are demonstrated based on graphene superlattices: an electromagnetic radiation shield with 97.5% effectiveness, a tunable far-infrared notch filter with 8.2 dB rejection ration, and a tunable terahertz linear polarizer with 9.5 dB extinction ratio. Most importantly, our demonstration opens an avenue for the realization of more complex 3-dimensional graphene photonic devices such as detectors and modulators and meta-material systems in both mid- and far-infrared wavelength ranges.

**Methods**

The growth and transfer of large area graphene were performed using the approaches reported by Li, *et al*[32]. The quartz substrates were spin-coated with 20 nm of commercially available organic buffer layer (NFC)[26] before the transfer. After graphene transfer, the sample was exposed to nitric acid vapor for 15 minutes to increase the doping[33]. Before the subsequent transfer, another buffer layer was spin-coated. This organic buffer layer minimizes the impact of the quartz substrate, leading to a reduced scattering width as discussed in the main text. Moreover, the coating also avoids reduction in doping concentration in the subsequent graphene transfer processes. The transfer processes were repeated until desirable layer numbers were achieved. We used electron beam lithography and oxygen plasma etching to define graphene disks and micro-ribbons. The pattered areas were always 3.6 mm wide by 3.6 mm long, much larger than the infrared light beam size used for measurements.

For the mid- and far- infrared measurements, we used a Nicolet-8700 FTIR spectrometer in combination with a liquid helium cooled silicon bolometer. All measurements were performed at room temperature in a nitrogen environment. We measured the transmission $T$ through the graphene area and a reference area without graphene on the same sample ($T_0$). The extinction ratio, $1-T/T_0$, was then obtained. The transmission spectra from the visible to ultra-violet were recorded using a Perkin Elmer UV-VIS spectrometer.

Plasmonic extinction spectra were fit using Eqs (1) and (2) in the main text. Drude weight $D$, plasmonic resonance frequency $\omega_p$ and the resonance width $\Gamma_p$ are three fitting parameters. The fitted Drude weight is in good agreement with the experimentally measured values from the Drude response in the unpatterned area. The fitted plasmonic resonance width $\Gamma_p$ is usually 10% larger than the Drude scattering width $\Gamma$ measured in the unpatterned area, probably due to the sample inhomogeneity and additional scattering at the edge. The fitted resonance frequency $\omega_p$ is consistent with Eq. (3). Because the organic buffer layer is very thin (20 nm) compared to the wavelength of interest (> 50 μm) and non-conductive, we neglect its impact on the dynamical conductivity.

**Acknowledgments**

The authors are grateful to B. Ek and J. Bucchignano for technical assistance and DARPA for partial financial support through the CERA program (contract FA8650-08-C-7838).

**Author contributions**

F.X., P. A., and H. Y. conceived the experiments, F. X. and H. Y fabricated the devices, H. Y. performed the measurements and data analysis, X. L. and B. C. grew the CVD

graphene, G. T. helped with doping, M. F. helped with the experimental setups, Y. W. and W. Z. participated in the sample fabrication and characterization. F. X. and H. Y. co-wrote the manuscript, P. A. provided suggestions and all the authors commented on the manuscript.

**Figure Captions**

**Figure 1: Fabrication of transparent graphene superlattices**

(a) Quartz is used as a transparent substrate and the fabrication involves three layer deposition steps and a single lithographic step: 1. Coating of the polymer buffer layer, 2. Deposition of wafer-scale graphene, 3. Doping of the graphene, and 4. Patterning of the superlattice into desirable structures. Here graphene superlattice disk arrays are shown. (b) A scanning electron micrograph (in false color) of a graphene superlattice micro-disk array arranged in a triangular lattice. Here $d$ = 2.6 μm and $a$ = 3 μm. (c) Relative transmission spectra of an as-prepared superlattice consisting of five graphene layers from near-IR to UV. The relative transmission in the visible is around 90% and an excitonic peak is observed at around 260 nm[17]. Inset: a photograph of the as-prepared superlattice before lithographic patterning.

**Figure 2: Electromagnetic wave shielding using as-prepared transparent graphene superlattices**

(a) Extinction in transmission, $1-T/T_0$, using a single layer of unpatterned graphene in far-infrared and terahertz wavelength range. Grey (red) squares: undoped graphene on quartz without (with) the polymer buffer layer underneath. Green square: doped graphene on quartz with polymer buffer layer. Solid lines are corresponding fitted curves. Inset: the schematic of the measurement. (b) The shielding effectiveness using the superlattices with 1, 2, 3, 4, and 5 layers of graphene, respectively. Solid lines are fitted curves. (c) Fitted Drude weight ($D_{Total}$) and scattering width ($\Gamma$) as a function of graphene layer number in the superlattice.

**Figure 3: Plasmons in patterned graphene superlattices at strong coupling limit**

(a) Extinction in transmission, $1-T/T_0$, in plasmonic superlattices with 1, 2, and 5 graphene layers. The graphene plasmonic superlattice is formed by patterning the layers into microdisks in a triangular lattice as shown in Fig. 1b. The solid lines are fitted curves. Here, $d = 4.4$ μm and $a = 9$ μm. Inset: the schematic view of the disk-disk coupling in two closely stacked graphene disks. (b) The normalized plasmonic resonance frequency versus the normalized total carrier density. Grey (red) squares: enhancement of the carrier density is realized using more graphene layers in superlattice case (using higher chemical doping in a single layer graphene case). Solid lines are power-law scaling curves. (c) The normalized dynamical conductivity at plasmonic resonance versus the normalized total carrier density. Grey (red) squares: enhancement of the at-resonance dynamical conductivity $\sigma_{Total}$ is realized using more graphene layers in superlattice case (or using higher chemical doping in a single layer graphene case). $\sigma_0$ is the at-resonance dynamical conductivity before doping in single layer graphene case, and is the at-resonance dynamical conductivity of an individual graphene layer in superlattice case. Solid lines are power-law scaling curves.

**Figure 4: Transparent far-infrared filters and terahertz polarizers**

(a) Extinction in transmission, $1-T/T_0$, in single layer graphene plasmonic superlattices with identical $d$ (4.4 µm) but different $a$ (4.8 and 9 µm, respectively). The resonance softens when smaller $a$ is used, due to graphene disk-disk interaction within the same layer. Inset: the schematic of the graphene disk-disk interaction. (b) Extinction spectra of tunable terahertz filters using superlattices with 5 graphene layers. The resonance frequency can be tuned by varying the diameter of the disks. In these filters, the lattice constant $a$ is always 400 nm larger than the disk diameter $d$. (c) Extinction spectra of a graphene superlattice polarizer for light polarizations along ($\theta = 0^0$) and perpendicular ($\theta = 90^0$) to the micro-ribbons. Lower left inset: the schematic of the polarizer. $\theta$ represents the angle between the E-field and the micro-ribbons. The width of the ribbons ($W$) is 2 µm and the ribbon-to-ribbon spacing ($S$) is 500 nm. Upper right inset: the extinction ($1-T/T_0$) at 40 cm$^{-1}$ as a function of light polarization, plotted in a polar coordinate system.

# Figure 1 Fabrication of transparent graphene superlattices

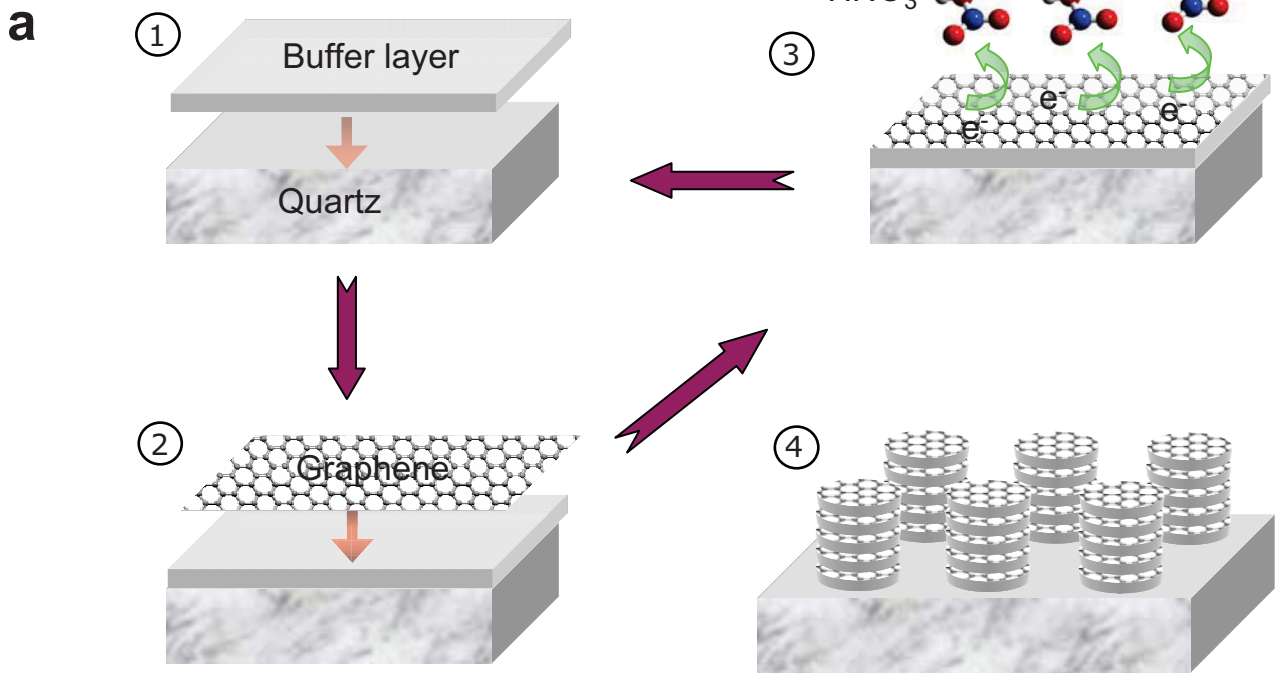

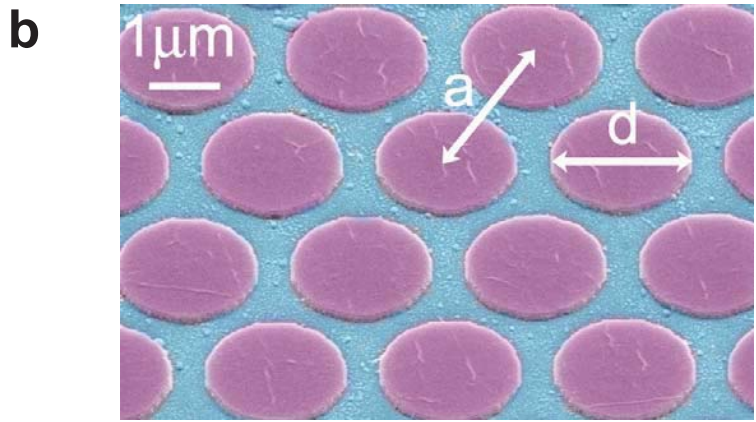

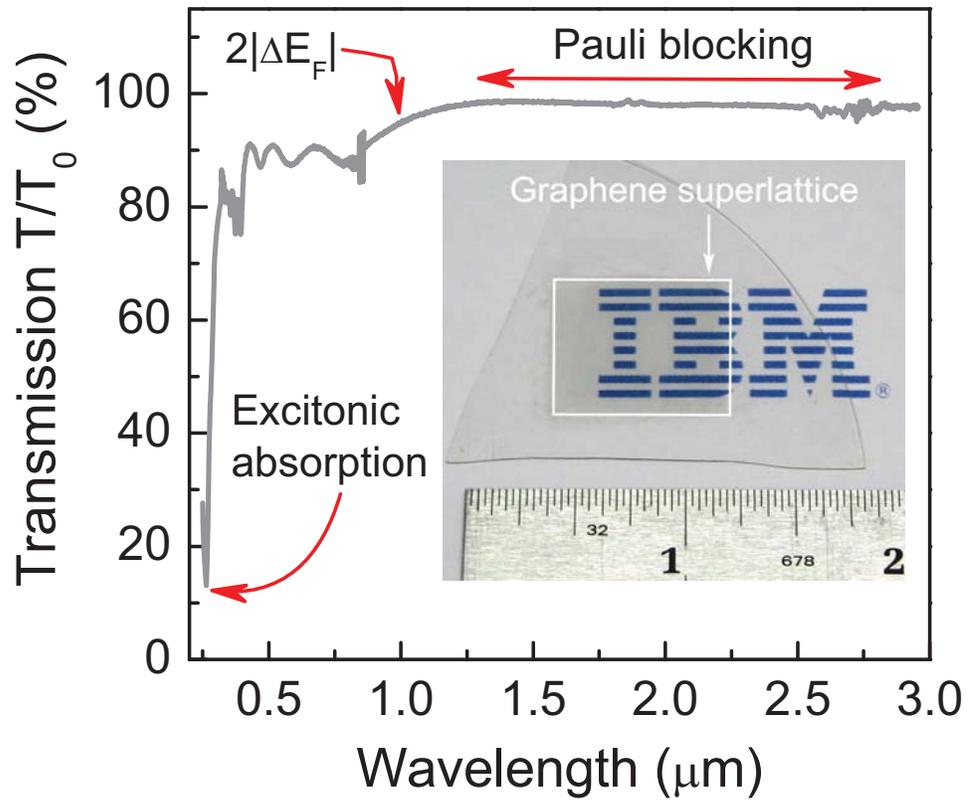

# Figure 2  Electromagnetic wave shielding using as-prepared graphene superlattices

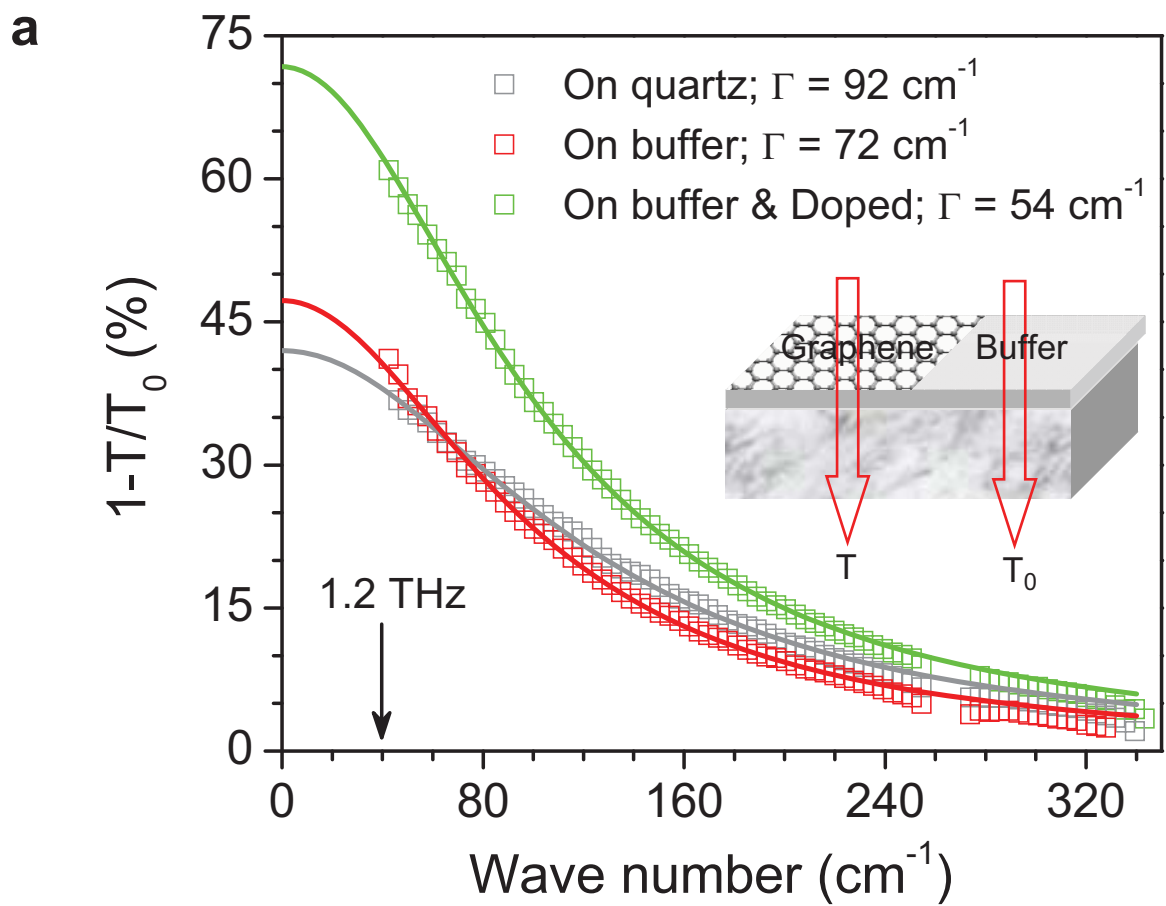

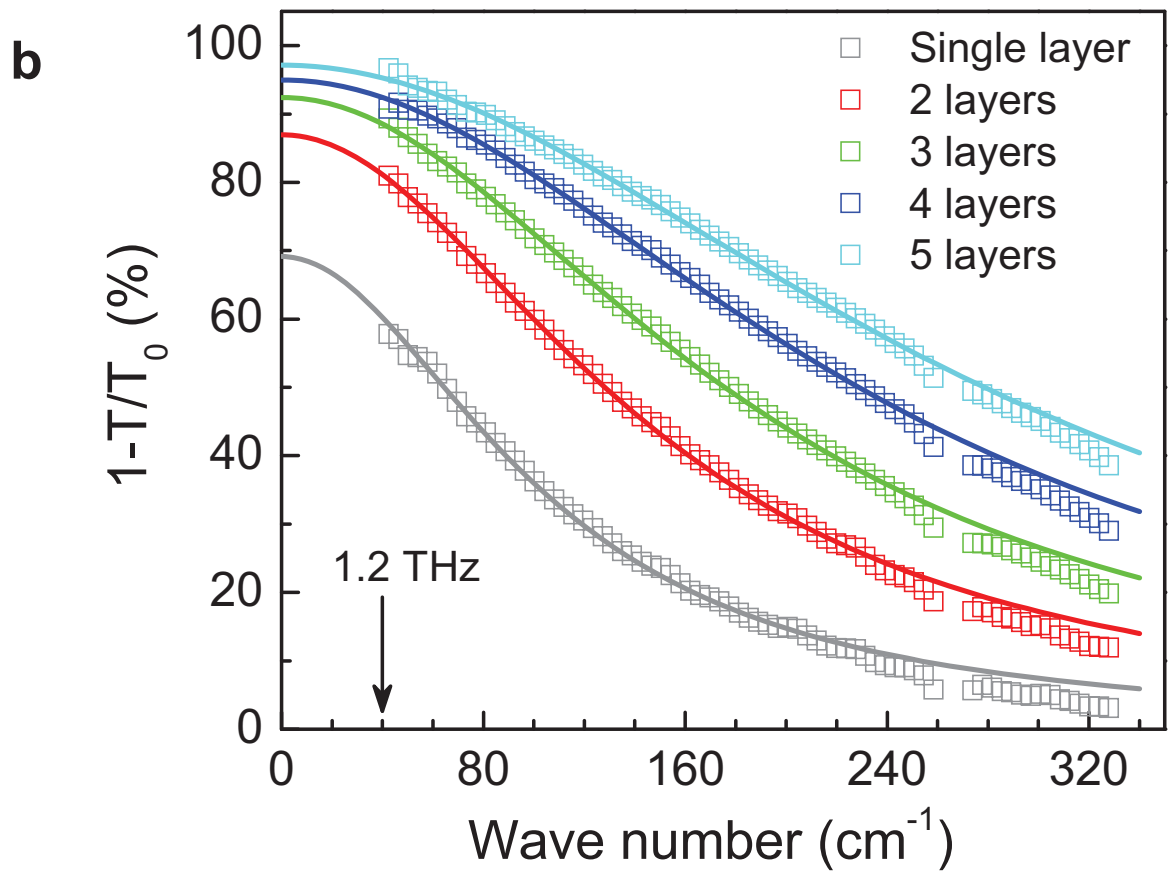

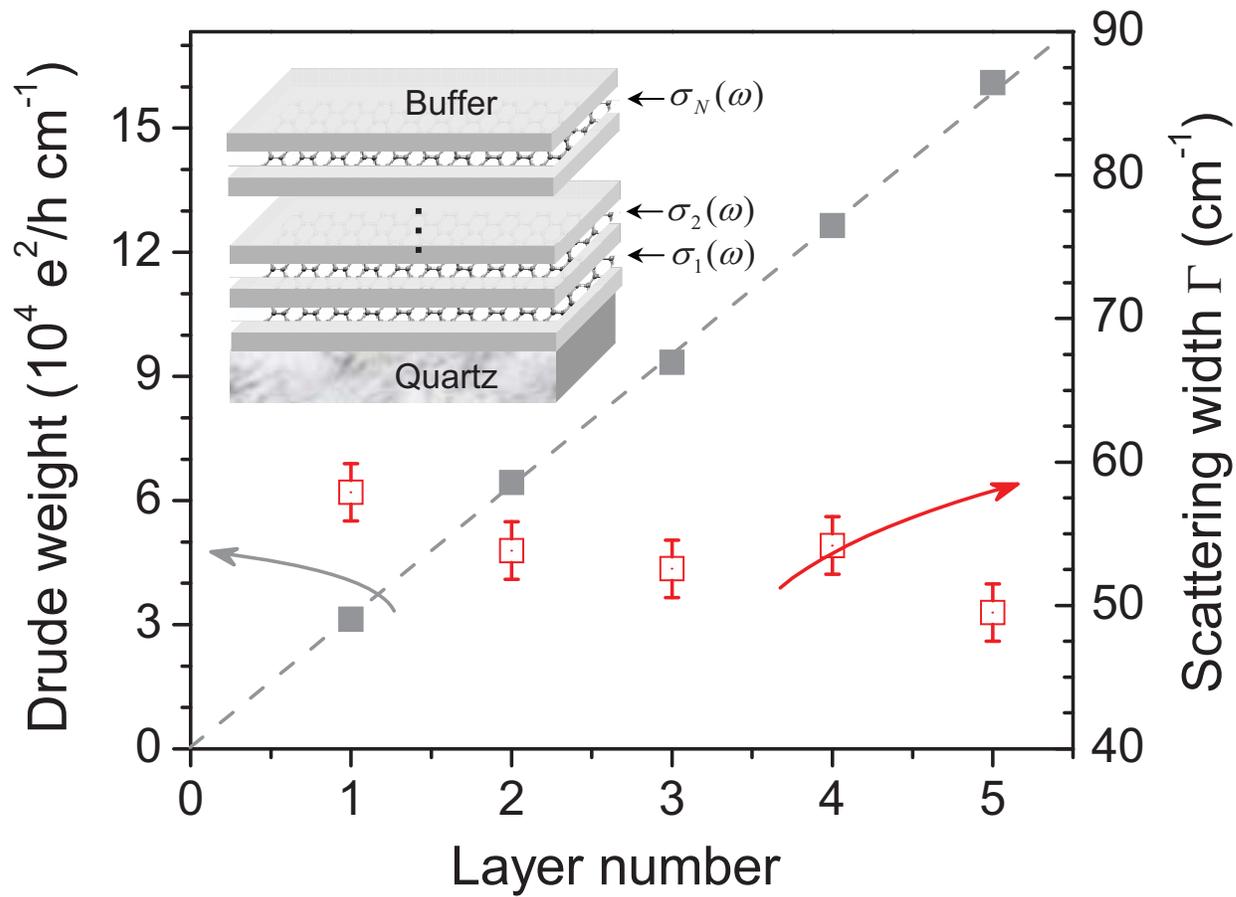

# Figure 3 Plasmons in graphene superlattice at strong coupling limit

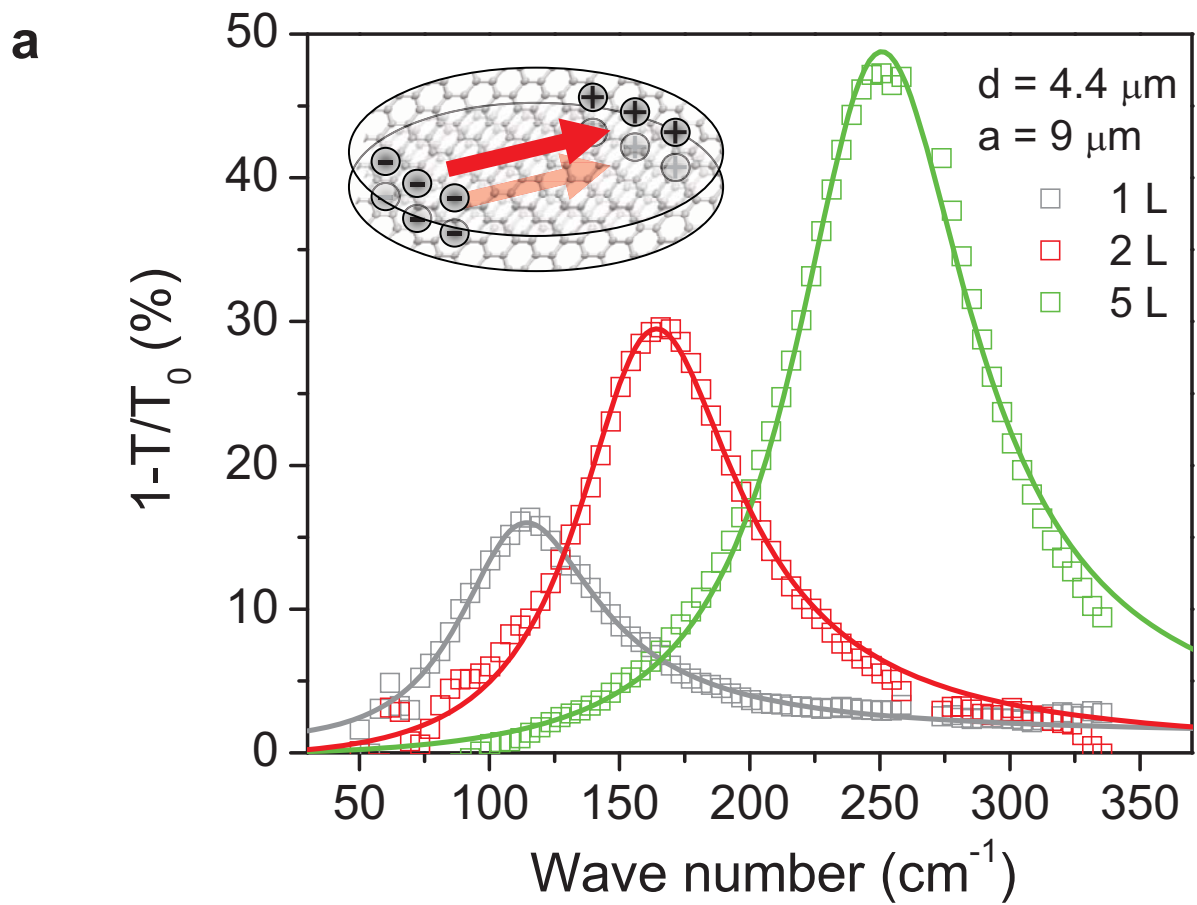

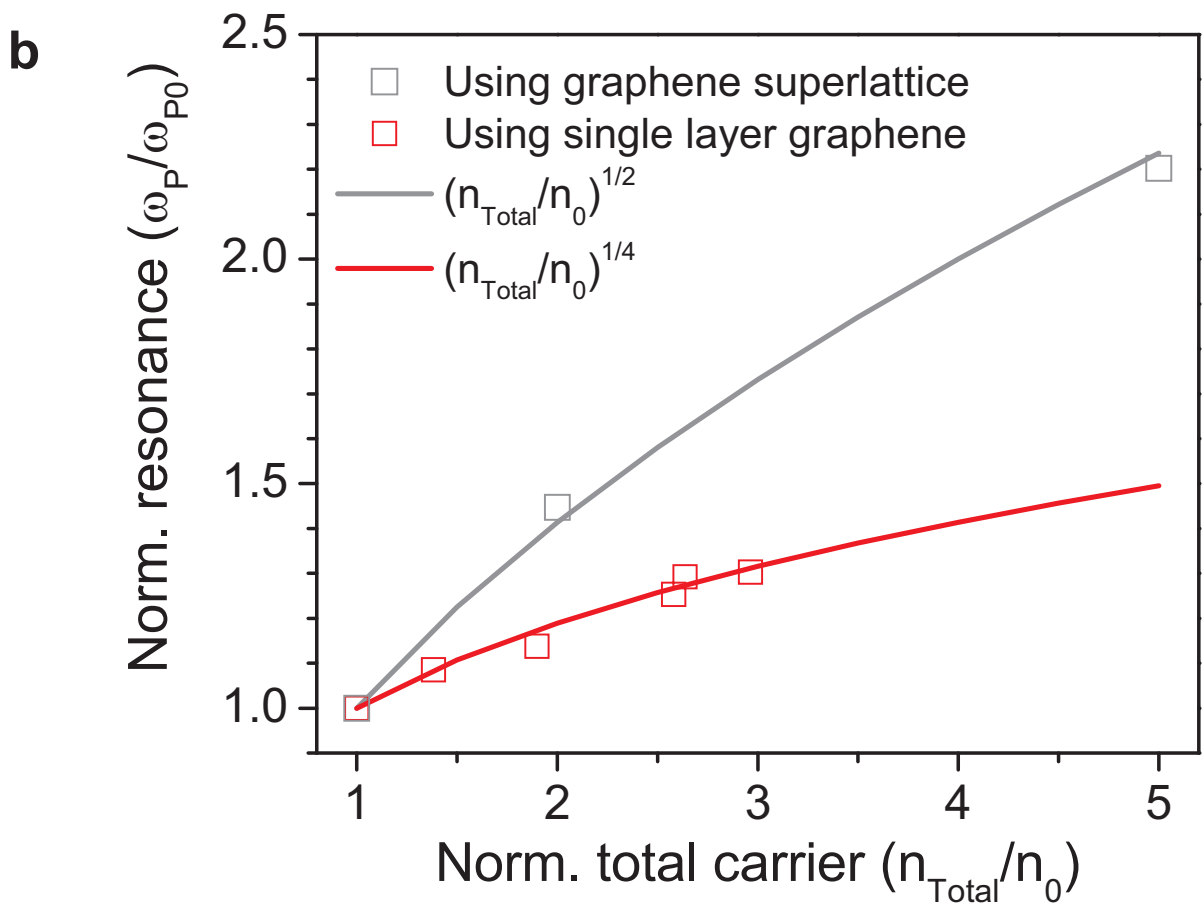

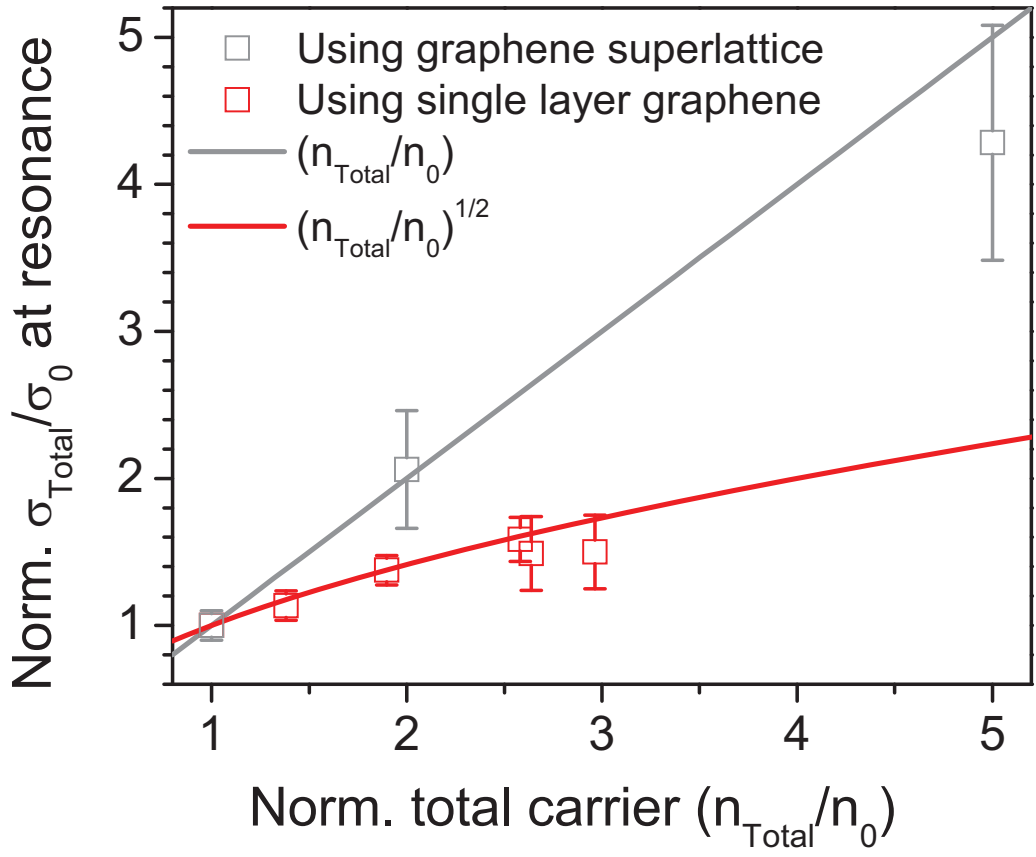

**Figure 4 Transparent far-infrared filters and terahertz polarizers**

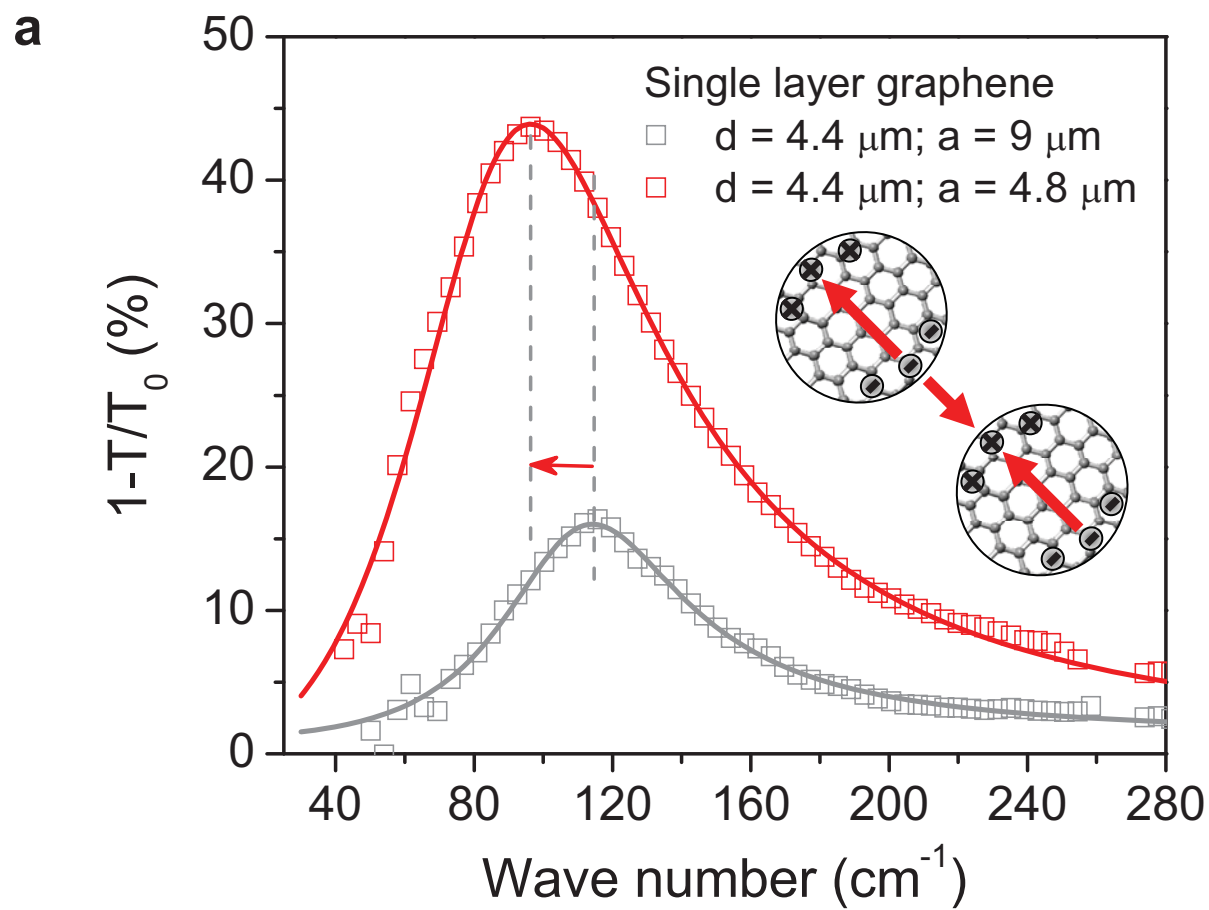

a

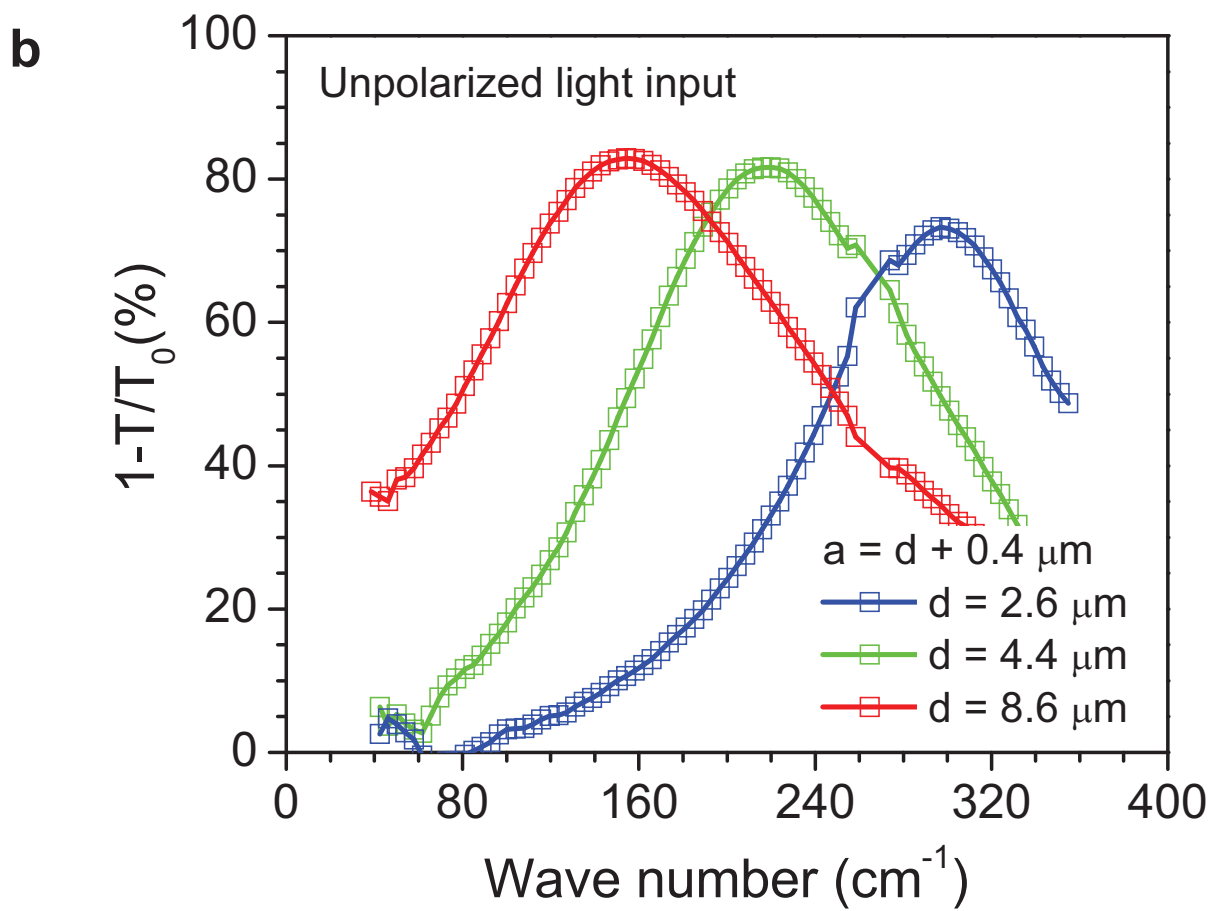

b

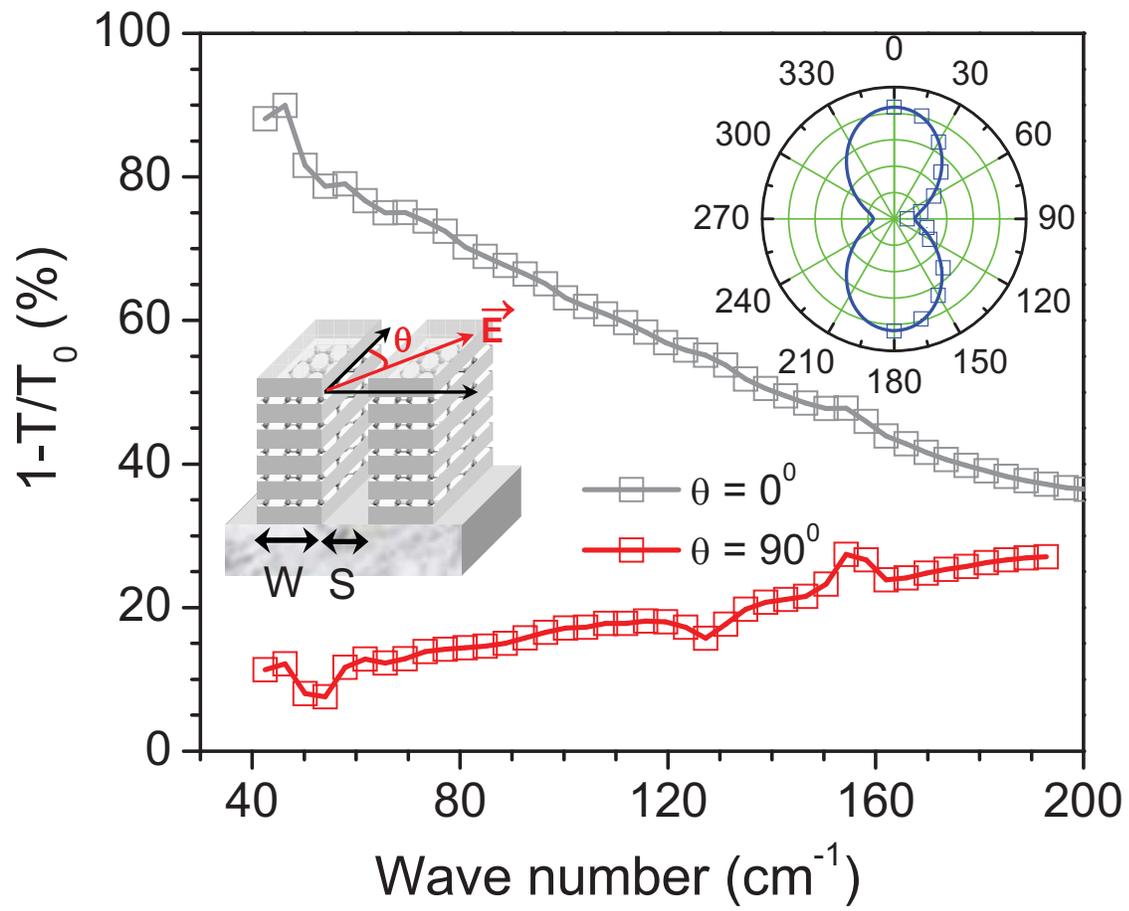

# Supplementary Information for "Manipulating infrared photons using plasmons in transparent graphene superlattices"

Hugen Yan, Xuesong Li, Bhupesh Chandra, George Tulevski, Yanqing Wu,

Marcus Freitag, Wenjuan Zhu, Phaedon Avouris[*], and Fengnian Xia[*]

*IBM Thomas J. Watson Research Center, Yorktown Heights, NY 10598*

**I. Determination of carrier density in graphene using spectroscopy**

Figure S1 shows the mid-IR extinction spectra ($1-T/T_0$) of a single layer graphene on quartz substrate. Before doping, the difference between the Fermi-level and the Dirac point energies, $\Delta E_F$, inferred from the extinction spectrum is about 350 meV. After doping, this value is typically increased to about 600 meV and can not be determined using our FTIR setup. However, it can be measured using another spectroscopy setup covering near-IR and visible wavelength range, as shown in Fig. 1c in the main text.

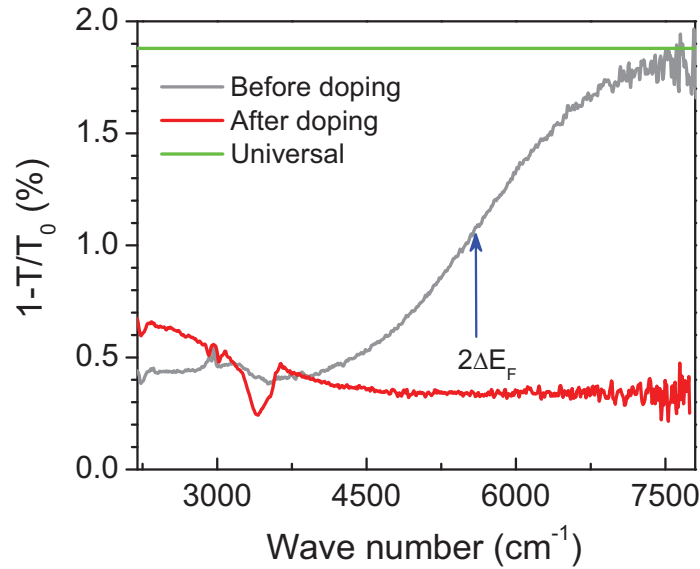

Figure S1. Mid-IR extinction spectra for CVD graphene on quartz before and after nitric acid doping. The vertical blue arrow indicates the estimated location of the onset of Pauli-blocking.

The universal extinction value of 1.9% for single layer graphene on quartz is also shown in Fig. S1, which is slightly smaller than 2.3% measured on suspended graphene, due to the slightly different boundary conditions for graphene on quartz. Most of our doped graphene samples were covered with polymer buffer layer. When we studied the impact of doping on plasmonic resonances in single layer graphene (Figs. 3b and 3c in the main text), the doped samples were intentionally left uncovered. By baking the exposed graphene sample in air at $170^0$C for 30 minutes, up to 70% in doping reduction can be achieved. Through controlling the acid exposure and baking times, we achieved different doping levels. From the measured far infrared Drude response of the single layer graphene, we can infer the DC sheet resistance by extrapolating it to zero frequency as shown in Figs. 2a and 2b in main text. Typical values for the as-prepared and chemically doped CVD graphene are 300Ω/□ and 130 Ω/□, respectively. The sheet resistance of the doped graphene reported here is comparable to the best results published previously[S1].

We can obtain the doping concentrations for graphene samples based on the Pauli blocking position in the extinction spectra from mid-IR to ultraviolet, as discussed above. Sometimes the doping concentrations were also verified using the Raman G modes. On the other hand, the scattering width $\Gamma$, as well as Drude weight $D$ can be inferred using the far-IR extinction spectra, as discussed in the main text. By comparing the doping concentration extracted from Pauli blocking position and the Drude weight obtained from far-IR extinction, we found that there was no Drude weight reduction for all the samples reported in this work, in contrast to previous observation[S2]. However, for some lower quality samples with larger scattering rate, we did observe Drude weight reduction, which also varies significantly from sample to sample. The normalized carrier concentration for the single layer graphene in Figs. 3b and 3c of the main text was obtained from the measurements of Drude weight on unpatterned graphene areas.

**II. Disk-disk interaction among disks in the same graphene layer**

In the main text, we show that stacking graphene disks vertically greatly enhances the dipole-dipole interaction (see inset of Fig. 3a in the main text), which stiffens the plasmonic resonance frequency. For graphene disk arrays arranged in triangular lattice, if the disk-disk edge distance is comparable to the disk diameter, disk-disk interaction among adjacent disks softens the resonance frequency (see Fig. 4a in the main text). Here we present more detailed studies on the disk-disk interaction among disks within the same graphene layer. Figure S2 shows the extinction spectra for three different single layer graphene disk arrays arranged in triangular lattice. They all have the same disk diameter $d$ of 0.8 μm but different lattice constant $a$. These disks are on $SiO_2$/Si substrate and they are not intentionally doped. With decreasing lattice constant $a$ and hence the disk-disk edge distance, the resonance frequency is softened. In the meantime, due to an increasing in filling factor $f$, the peak amplitude increases drastically.

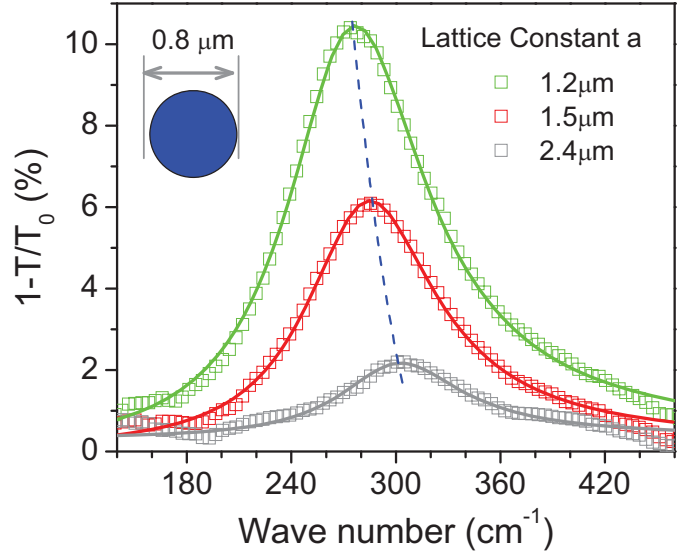

Figure S2. Measured extinction spectra of graphene disk arrays with diameter of 0.8 μm and different lattice constants. The peak shift is indicated by the dashed blue line. Solid lines are fitting curves based on Equations (1) and (2) in the main text.

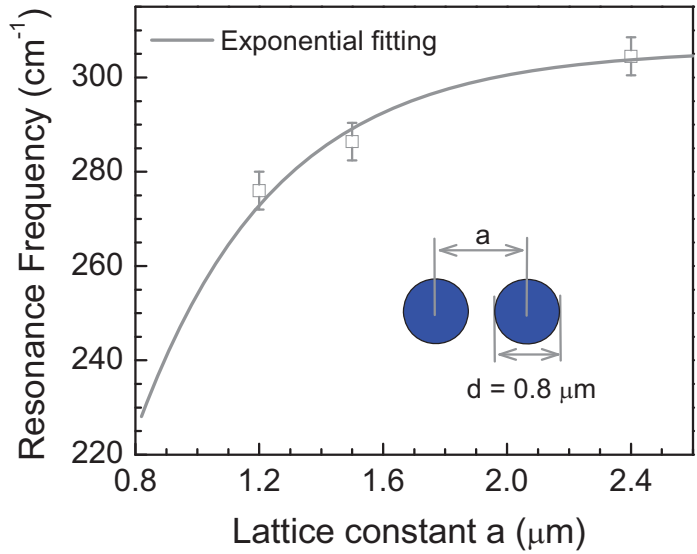

Figure S3 Resonance frequency as a function of the lattice constant *a*.

We plot the resonance frequencies in Fig. S2 as a function of the lattice constant *a* in Fig. S3. The grey line is a phenomenological exponential decay fitting with a decay constant of 0.45 μm[S3]. From the fitting, we conclude that if *a* is larger than 2 μm (2.5 times of the diameter of the disk), the disk-disk interaction effect within the same graphene layer is negligible. This is in good agreement with previous studies of interaction in metallic nanoparticles[S3].

*Email: fxia@us.ibm.com (F.X.) and avouris@us.ibm.com (P.A.)